\documentclass{article}%
\usepackage{amssymb}
\usepackage{amsmath}
\usepackage{amsfonts}
\usepackage{bm}
\usepackage{braket}
\usepackage{graphicx}%
\usepackage{latexsym}
\providecommand{\U}[1]{\protect\rule{.1in}{.1in}}

\begin{document}
\begin{titlepage}

\ \\
\begin{center}
\LARGE
{\bf
A Fundamental  Upper Bound\\
for Signal to Noise Ratio\\
of Quantum Detectors
}
\end{center}

\begin{center}
\large{Ryota Katsube, Masahiro Hotta, and Koji Yamaguchi }\\
{\it
Graduate School of Science, Tohoku University, Sendai, 980-8578, Japan}
\end{center}
\ \\
\begin{abstract}
Quantum fluctuations yield inevitable noises in quantum detection.
We derive an upper bound of signal to noise ratio for arbitrary quantum detection described by trace-class operators with discrete spectra.
The bound is independent of observables to be detected and is computed by  quantum fidelity of two initial quantum states.
We provide applications of the upper bound.
\end{abstract}
\end{titlepage}

\ \ 

\section{Introduction}

\bigskip

\quad Any quantum computation and quantum communication requires quantum detection of
observables in read-out procedures of the protocols. Efficiency analysis of quantum detectors is crucial in developing
a technology. In order to analyze this, there are several methods which we can consider. One of them  is the optimization of error probability in state discrimination measurements \cite{Helstrom,Holevo}. Another one is the quantum state tomography which was studied in 
\cite{Vogel,Leonhardt,Fujiwara2,Matsumoto2}.
The Quantum process tomography have been investigated in
\cite{Turchette,Chuang,Poyatos}. The analyses using quantum Fisher information are also available in  \cite{Fujiwara,Imai,Amari,Miller,Cafaro}. Fundamental restrictions of quantum measurements appear due to uncertainty relations of physical quantities in measurement errors, disturbances and standard deviations \cite{Ozawa2}. It is also known that the quantum detection efficiency is restricted by conservation laws of physical observables \cite{Wigner,Araki,Ozawa,Karasawa1,Karasawa2,Tajima}.\\
\quad In this paper, we focus on the signal to noise ratio (SNR) of quantum measurements. The SNR is crucial since it is  directly related to realistic measurements. Especially in quantum optics,
analysis of quantum SNR has been performed in earlier studies for
individual subjects, including squeezed states \cite{1,2,3}, weak
measurements \cite{4,5,6,David,Lee,Susa}, multimode spatial entanglement
detection\cite{7}, electron-multiplying CCD camera \cite{8}, heralded linear
amplifier \cite{9}, and correlation plenoptic imaging \cite{10}.
Inevitable noises occur due to quantum fluctuations of the
observables and may dominate in the quantum regime. We
provide a universal upper bound of quantum SNR for arbitrary quantum
detection described  by  trace-class  operators  with  discrete  spectra. \\
\quad The quantum detectors have the following 
setup: a target system $S$ carring relevant information in a quantum state
interacts with a measurement device system $D$ in finite duration. The initial
state $\rho_{D,0}$ of $D$ is independent of the information. The
information of $S$ is extracted by measuring an observable $\hat{A}_{D}$ of
$D$. Let us consider two initial quantum states $\rho_{S}(1)$ and $\rho
_{S}(2)$\ of $S$. The difference between the two states provides the information of
our interest to be detected. After the interaction between $S$ and $D$, the
final state of $D$ is computed as
\begin{equation}
\rho_{D}(s)=\operatorname*{Tr}_{S}\left[  U\left(  \rho_{S}(s)\otimes\rho
_{D,0}\right)  U^{\dag}\right]  , \label{31}%
\end{equation}
where \thinspace$s=1,2$ and $U$ is the unitary time evolution operator of the $S+D$
system. The above setup may look specific. However this corresponds to   the most general measurement due to the Stinespring factorization theorem regarding the representation of positive operator valued measures (POVMs)  by projection valued measures (PVMs) on an auxiliary quantum system \cite{naimark,ozawa-povm}. \\
\quad We define the quantum signal for $\hat{A}_{D}$ as
\begin{equation}
S_{D}=\left\vert \operatorname*{Tr}_{D}\left[  \hat{A}_{D}\rho_{D}(1)\right]
-\operatorname*{Tr}_{D}\left[  \hat{A}_{D}\rho_{D}(2)\right]  \right\vert,
\label{1}%
\end{equation}
and its quantum noise as%
\begin{equation}
N_{D}=\sqrt{\operatorname*{Tr}_{D}\left[  \left(  \hat{A}_{D}%
-\operatorname*{Tr}_{D}\left[  \hat{A}_{D}\rho_{D}(1)\right]  \right)
^{2}\rho_{D}(1)\right]  }+\sqrt{\operatorname*{Tr}_{D}\left[  \left(  \hat
{A}_{D}-\operatorname*{Tr}_{D}\left[  \hat{A}_{D}\rho_{D}(2)\right]  \right)
^{2}\rho_{D}(2)\right]  }. \label{2}%
\end{equation}
\quad We state here the main result of our paper (which we prove in the next section):
The upper bound of the SNR is given by
\begin{equation}
\frac{S_{D}}{N_{D}}\leq\frac{\sqrt{1-F\left(  \rho_{S}(1),\rho_{S}(2)\right)
^{2}}}{1-\sqrt{1-F\left(  \rho_{S}(1),\rho_{S}(2)\right)  ^{2}}}, \label{3}%
\end{equation}
where $F\left(  \rho_{S}(1),\rho_{S}(2)\right)  $ is the quantum fidelity of
$\rho_{S}(1)$ and $\rho_{S}(2)$ which is defined by%

\[
F\left(  \rho_{S}(1),\rho_{S}(2)\right)  =\operatorname*{Tr}\left[
\sqrt{\sqrt{\rho_{S}(1)}\rho_{S}(2)\sqrt{\rho_{S}(1)}}\right]  .
\]
Our result includes the standard SNR case as follows. When the initial state of $D$ is
given by $\rho_{D,0}=|a\rangle\langle a|\,$\ where $|a\rangle$ is an
eigenstate of the observable $\hat{A}_{D}$ such that $\hat{A}_{D}|a\rangle=a|a\rangle$ and
$U\left(  \rho_{S}(2)\otimes\rho_{D,0}\right)  U^{\dag}=\rho_{S}%
(2)\otimes\rho_{D,0}$ holds, $a=\operatorname*{Tr}_{D}\left[  \hat{A}_{D}%
\rho_{D}(2)\right]  $ is interpreted as the initial value of $\hat{A}_{D}$
and $\rho_{S}(2)$ as a no-signal state. Then $N_{D}$ becomes the standard
noise of $\hat{A}_{D}$ for $\rho_{D}(1)$ given by
\[
N_{D}=\sqrt{\operatorname*{Tr}_{D}\left[  \left(  \hat{A}_{D}%
-\operatorname*{Tr}_{D}\left[  \hat{A}_{D}\rho_{D}(1)\right]  \right)
^{2}\rho_{D}(1)\right]  }.
\]
It should be stressed that the upper bound of the quantum SNR in eq. (\ref{3})
does not depend on the observable $\hat{A}_{D}$ and is computed only by use
of the initial quantum fildelity $F\left(  \rho_{S}(1),\rho_{S}(2)\right)  $ of $S$.
\\
\quad This paper is organized as follows. In section 2, we prove the upper bound of
the quantum SNR given in eq. (\ref{3}). In section 3, we discuss applications of our
results. In section 4,  a summary is given. We adopt the natural units $c=\hbar=1$
in this paper.

\bigskip

\section{Derivation of Universal Upper Bound of Quantum SNR}

\bigskip

\quad In this section we derive the upper bound for the SNR in eq. (\ref{3}). The dimension of the
Hilbert space of $D$ is denoted by $d$. We assume that the observable $\hat{A}_D$ has a finite discrete spectrum. The eigenvalues of $\hat{A}_{D}$ are
denoted by $\{a_{i}\}$ and they satisfy $a_{1}\leq a_{2}\leq\cdots\leq a_{d}$. Let
us denote a normalized eigenvector associated with $a_{i}$ by $|a_{i}\rangle$.
The spectral decomposition of $\hat{A}_{D}$ is given by $\hat{A}_{D}%
=\sum_{i=1}^{d}a_{i}|a_{i}\rangle\langle a_{i}|$. For a quantum state
$\rho_{D}(1)$ of $D$, the probability of observing $a_{i}$ is computed as
$p_{i}=\langle a_{i}|\rho_{D}(1)|a_{i}\rangle$. Similarly, for a quantum state $\rho_{D}(2)$ of $D$, $q_{i}$ is given by$\langle
a_{i}|\rho_{D}(2)|a_{i}\rangle$.
The fidelity $F(\{p_i\},\{q_i\})$ and  Bures distance $L_{B}(\{p_i\},\{q_i\})$ between the above classical
probability distributions of $\{p_{i}\}$ and $\{q_{i}\}$ are defined as
\[
F(\{p_i\},\{q_i\})=\sum_{i}\sqrt{p_{i}q_{i}},
\]
and%
\[
L_{B}(\{p_i\},\{q_i\})=\sqrt{1-F(\{p_i\},\{q_i\})}=\sqrt{1-\sum_{i}\sqrt{p_{i}q_{i}}} .
\]
\\
\quad We first prove the following Lemma: The upper bound of the quantum signal $S_D$ is given by
\begin{equation}
S_{D}\leq\sqrt{2-L_{B}(\rho_{D}(1),\rho_{D}(2))^{2}}L_{B}(\rho_{D}(1),\rho_{D}(2))\left[
\sqrt{\mathrm{Tr}_{D}[\hat{A}_{D}^{2}\rho_{D}(1)]}+\sqrt{\mathrm{Tr}_{D}[\hat
{A}_{D}^{2}\rho_{D}(2)]}\right]  .  \label{5}
\end{equation} \\
Proof: Let us introduce a real vector $\overrightarrow{X}$ whose $i$th component is
given by $\sqrt{p_{i}}$. Similarly, $\overrightarrow{Y}$ is defined as a real
vector whose $i$th component is $\sqrt{q_{i}}$. Note that the
following relations hold using inner products of the vectors:
\begin{align}
\overrightarrow{X}\cdot\overrightarrow{Y}  &  =F(\{p_i\},\{q_i\}),\nonumber\\
\overrightarrow{X}\cdot\overrightarrow{X}  &  =1,\nonumber\\
\overrightarrow{Y}\cdot\overrightarrow{Y}  &  =1.\nonumber
\end{align}
In terms of the vectors $\overrightarrow{X}$ and $\overrightarrow{Y}$, $S_{D}$ in eq. (\ref{1}) is written as
\begin{equation}
S_{D}=\left\vert \sum_{i}a_{i}(p_{i}-q_{i})\right\vert =|\overrightarrow
{X}\cdot A_D \overrightarrow{X}-\overrightarrow{Y}\cdot A_D\overrightarrow
{Y}|,\nonumber
\end{equation}
where $A_D$ is a $d$-dim matrix given by $A_D=\left[
a_{i}\delta_{ij}\right]  $.
By straightforward calculation, the following relation also holds:%
\[
\sqrt{\operatorname*{Tr}_{D}\left[  \hat{A}_{D}^{2}\rho_{D}(1)\right]  }%
+\sqrt{\operatorname*{Tr}_{D}\left[  \hat{A}_{D}^{2}\rho_{D}(2)\right]  }%
=\sqrt{\sum_{i}a_{i}^{2}p_{i}}+\sqrt{\sum_{i}a_{i}^{2}q_{i}}=\sqrt
{\overrightarrow{X}\cdot A_D^{2}\overrightarrow{X}}+\sqrt{\overrightarrow
{Y}\cdot A_D^{2}\overrightarrow{Y}}.
\]
If both $\overrightarrow{X}$ and $\overrightarrow
{Y}$ are eigenvectors associated with eigenvalue $0$ of $A_D$, the problem
becomes trivial and the relation in eq. (\ref{5}) is satisfied. In the
following, let us consider other nontrivial cases where one of $\overrightarrow{X}$
and $\overrightarrow{Y}$ is not an eigenvector associated with eigenvalue $0$
of $A_D$. In this case we are able to define $f_{A_D}(\overrightarrow{X}%
,\overrightarrow{Y})$ as follows:
\begin{equation}
f_{A_D}(\overrightarrow{X},\overrightarrow{Y})=\left(  \frac{\overrightarrow
{X}\cdot A_D \overrightarrow{X}-\overrightarrow{Y}\cdot A_D\overrightarrow{Y}%
}{\sqrt{\overrightarrow{X}\cdot A_D^{2}\overrightarrow{X}}+\sqrt{\overrightarrow
{Y}\cdot A_D^{2}\overrightarrow{Y}}}\right)  ^{2}.\nonumber
\end{equation}
Note that the inner products $\overrightarrow{X}\cdot A_D \overrightarrow
{X},~\vec{Y}\cdot A_D \overrightarrow{Y},\overrightarrow{X}\cdot A_D^{2}
\overrightarrow{X}\,\ $\ and $\overrightarrow{Y}\cdot A_D^{2}\overrightarrow{Y}$
are invariant under the following coordinate transformation:
\[
\overrightarrow{X}^{\prime}=R\overrightarrow{X},\quad\overrightarrow
{Y}^{\prime}=R\overrightarrow{Y},\quad A^{\prime}=RA_DR^{\mathrm{T}},
\]
where $R$ is an arbitrary orthogonal matrix satisfying $RR^{\mathrm{T}}=I$.
Thus $f_{A_D}(\overrightarrow{X},\overrightarrow{Y})$ is also invariant. Using
the above symmetry of $f_{A_D}(\overrightarrow{X},\overrightarrow{Y})$, without
loss of generality, we are able to fix $R$ to a specific matrix so that only
the first two components of two vectors $\overrightarrow{X}$ and
$\overrightarrow{Y}$ are nonvanishing as
\begin{eqnarray}
\overrightarrow{X}^{\prime}=\left[
\begin{array}
[c]{ccccc}%
x_{1} & x_{2} & 0 & \cdots & 0
\end{array}
\right]  ^{\mathrm{T}},\quad\overrightarrow{Y}^{\prime}=\left[
\begin{array}
[c]{ccccc}%
y_{1} & y_{2} & 0 & \cdots & 0
\end{array}
\right]  ^{\mathrm{T}}. \label{twovectors}
\end{eqnarray}
Let us consider a two-dimensional subspace spanned by $\overrightarrow{X}$ and $\overrightarrow{Y}$. Suppose that the two unit vectors  $\overrightarrow{e_1}$ and $\overrightarrow{e_2}$ are orthogonal to each other in this subspace. Then $\overrightarrow{X} = x_1 \overrightarrow{e_1} + x_2 \overrightarrow{e_2}$ and $\overrightarrow{Y} = y_1 \overrightarrow{e_1} + y_2 \overrightarrow{e_2}$. When we take $ R= [\overrightarrow{e_1}\quad  \overrightarrow{e_2} \quad \cdots \quad \overrightarrow{e_d}]^{T} $, where $\{\overrightarrow{e_i}\}_{i=1}^d$ is a basis in the total space, we are able to get the  above two vectors in eq. (\ref{twovectors}). 
Here, we define two-dimensional real vectors as follows:
\[
\overrightarrow{x}=\left[
\begin{array}
[c]{c}%
x_{1}\\
x_{2}%
\end{array}
\right]  ,\quad\overrightarrow{y}=\left[
\begin{array}
[c]{c}%
y_{1}\\
y_{2}%
\end{array}
\right]  .
\]
Note that $\overrightarrow{x}\cdot\overrightarrow{y}=F(\{p_i\},\{q_i\})$ is satisfied. Let
us define a two-dimensional matrix $B$ to be a submatrix of $A^{\prime}$ such that
\begin{equation}
B=\left[
\begin{array}
[c]{cc}%
A_{11}^{\prime} & A_{12}^{\prime}\\
A_{21}^{\prime} & A_{22}^{\prime}%
\end{array}
\right]  .\nonumber
\end{equation}
Then we obtain
\begin{equation}
s_{B}(x,y)=\left(  \frac{\overrightarrow{x}\cdot B\overrightarrow
{x}-\overrightarrow{y}\cdot B\overrightarrow{y}}{\sqrt{\overrightarrow{x}\cdot
B^{2}\overrightarrow{x}}+\sqrt{\overrightarrow{y}\cdot B^{2}\overrightarrow
{y}}}\right)  ^{2}=f_{A^{\prime}}(\overrightarrow{X}^{\prime},\overrightarrow
{Y}^{\prime}).
\end{equation}
Since $A_D$ is a real symmetric matrix, $B$ is also a real symmetric matrix.
The eigenvalues of $B$ are denoted by $b_{1}$ and $b_{2}$ satisfying $|b_{1}%
|\geq|b_{2}|$. Note that a trivial scale invariance of $s_{B}(x,y)$,
$s_{cB}(x,y)=s_{B}(x,y)$ holds, where $c$ is an arbitrary real number. Therefore it can be assumed that the matrix $B$ has
eigenvalues $1$ and $b=\frac{b_{2}}{b_{1}}$, where $\left\vert b\right\vert
\leq1$. The corresponding eigenvector for the eigenvalue $1$ is denoted by
$\overrightarrow{u_{1}}$ and the corresponding eigenvector for the eigenvalue
$b$ is $\overrightarrow{u_{2}}$, respectively. A spectral decomposition of $B$ is
given by $B=\overrightarrow{u_{1}}\overrightarrow{u_{1}}^{T}+b\overrightarrow
{u_{2}}\overrightarrow{u_{2}}^{T}$. Defining $\sqrt{P}=|\overrightarrow{u_1}\cdot \overrightarrow{x}|$ and $\sqrt{Q}=|\overrightarrow{u_1}\cdot \overrightarrow{y}|$, 
the function $s_{B}(x,y)$ is represented by a function \ $g(b,P,Q)$
of $b$, $P$ and $Q$ as
\begin{equation}
s_{B}(x,y)=g(b,P,Q)=\frac{(1-b)^{2}(P-Q)^{2}}{\left(  \sqrt{P+b^{2}%
(1-P)}+\sqrt{Q+b^{2}(1-Q)}\right)  ^{2}}. \label{eq:g}%
\end{equation}
To find the maximum of $g(b,P,Q)$ for fixed $P$
and $Q$, we vary the value of $b$. We solve the following equation:
\begin{align}
&  \frac{\partial g(b,P,Q)}{\partial b}\nonumber \\
&  =\frac{-2(1-b)(P-Q)^{2}}{\left(  \sqrt{P+b^{2}(1-P)}+\sqrt{Q+b^{2}%
(1-Q)}\right)  ^{3}}\left[  \frac{P+b(1-P)}{\sqrt{P+b^{2}(1-P)}}%
+\frac{Q+b(1-Q)}{\sqrt{Q+b^{2}(1-Q)}}\right] \nonumber \\
&  =0. \label{derivative}
\end{align}
\quad It is easy to check that the above equation has a trivial solution $P=Q$, 
which provides the minimum and we do not consider it further. Thus we focus on the case with $P\neq Q$ later.
\begin{enumerate}
\item $P\neq1$ and $Q\neq1$:\\
 When $P\neq1$ and $Q\neq1$, the nontrivial
solution of the above equation is given by
\begin{equation}
b=b^{\ast}=-\sqrt{\frac{PQ}{(1-P)(1-Q)}}.\label{attain}
\end{equation}
\begin{enumerate}
\item $P+Q \leq 1$: \\
In this case  $|b^{\ast}|\leq1$ holds and  $g(b,P,Q)$ takes the maximum $g(b^{\ast
},P,Q)=1-(\sqrt{PQ}+\sqrt{(1-P)(1-Q)})^{2}$ at $b=b^{\ast}$.
 Since the following inequality is satisfied:
\begin{eqnarray}
\sqrt{PQ}+\sqrt{(1-P)(1-Q)}\geq F(\{p_i\},\{q_i\}), \label{eq2}
\end{eqnarray}
we get 
\begin{equation}
g(b,P,Q)=\left(  \frac{\sum_{i}a_{i}(p_{i}-q_{i})}{\sqrt{\sum_{i}a_{i}%
^{2}p_{i}}+\sqrt{\sum_{i}a_{i}^{2}q_{i}}}\right)  ^{2}\leq1-F(\{p_i\},\{q_i\})^{2}
\label{eq:classical}.
\end{equation}

\item  $P+Q>1$: \\
It is stressed that the relation in eq. (\ref{eq:classical}) generally holds
even in the case that $P+Q>1$ since $g(b,P,Q)$ monotonically decreases in the range
$-1\leq b\leq1$ \ and $g(b,P,Q)$ takes the maximum value $(P-Q)^{2}$ at
$b=-1$.
\end{enumerate}
\item $P=1$ or $Q=1$:\\
Even when $P=1$ or $Q=1$, eq. (\ref{eq:classical}) trivially holds because $g(b,P,Q)$ monotonically decreases in the range
$-1\leq b\leq1$.\\
\end{enumerate}
\quad From the above computation, the relation holds for a fixed basis of $\hat
{A}_{D}$. Let us choose an arbitrary basis $\{|a_{i}\rangle\}$. The
classical fidelity $F(\{p_i\},\{q_i\})$ is greater than or equal to the quantum fidelity
$F(\rho_{D}(1),\rho_{D}(2))$ for any $\{|a_{i}\rangle\}$. Thus we obtain
\begin{equation}
1-F(\{p_i\},\{q_i\})^{2}\leq1-F(\rho_{D}(1),\rho_{D}(2))^{2}=\left[2-L_{B}\left(\rho_{D}(1),\rho_{D}(2)%
\right)^{2}\right]L_{B}(\rho_{D}(1),\rho_{D}(2))^{2}. \label{eq:cq}%
\end{equation}
From eq. (\ref{eq:classical}) and eq. (\ref{eq:cq}), it is possible to derive
\begin{equation}
\left(  \frac{\sum_{i}a_{i}(p_{i}-q_{i})}{\sqrt{\sum_{i}a_{i}^{2}p_{i}}%
+\sqrt{\sum_{i}a_{i}^{2}q_{i}}}\right)  ^{2}\leq(2-L_{B}(\rho_{D}(1),\rho
_{D}(2))^{2})L_{B}(\rho_{D}(1),\rho_{D}(2))^{2}\nonumber ,
\end{equation}
for arbitrary $\hat{A}_{D}$. This yields
\begin{equation}
S_{D}\leq\sqrt{2-L_{B}(\rho_{D}(1),\rho_{D}(2))^{2}}L_{B}(\rho_{D}(1),\rho
_{D}(2))\left[  \sqrt{\mathrm{Tr}[\hat{A}_{D}^{2}\rho_{D}(1)]}+\sqrt
{\mathrm{Tr}[\hat{A}_{D}^{2}\rho_{D}(2)]}\right]  . \label{eq:eq1}%
\end{equation}
Thus we have derived the lemma (\ref{5}). \\
\quad Next we consider a constant shift of the origin of eigenvalues of $\hat{A}_{D}$ by
$\alpha$ and define $\tilde{A}_{D}=\hat{A}_{D}-\alpha I$. This generates a
tighter inequality than (\ref{eq:eq1})  by optimizing $h(\alpha
)=\sqrt{\mathrm{Tr}[\tilde{A}_D^{2}\rho_{D}(1)]}+\sqrt{\mathrm{Tr}[\tilde{A}_D
^{2}\rho_{D}(2)]}$. The function $h(\alpha)$ attains the minimum at
\[
\alpha^{\ast}=\frac{\mathrm{Tr}[\hat{A}_{D}\rho_{D}(1)]\delta_{\hat{A}_D}(\rho
_{D}(2))+\mathrm{Tr}[\hat{A}_{D}\rho_{D}(2)]\delta_{\hat{A}_D}(\rho_{D}(1))}%
{\delta_{\hat{A}_D}(\rho_{D}(1))+\delta_{\hat{A}_D}(\rho_{D}(2))},
\]
where $\delta_{\hat{A}_D}(\rho)=\sqrt{\mathrm{Tr}[\hat{A}_{D}^{2}\rho]-\left(
\mathrm{Tr}[\hat{A}_{D}\rho]\right)  ^{2}}$.

Let us take $\alpha=\mathrm{Tr}[\hat{A}_{D}\rho_{D}(1)]$. Then the following
relations are obtained:
\begin{align*}
S_{D}  &  =\left\vert \mathrm{Tr}[\tilde{A}_D\rho_{D}(1)]-\mathrm{Tr}[\tilde
{A}_D\rho_{D}(2)]\right\vert \\
&  \leq\sqrt{2-L_{B}(\rho_{D}(1),\rho_{D}(2))^{2}}L_{B}(\rho_{D}(1),\rho
_{D}(2))h(\mathrm{Tr}[\hat{A}_{D}\rho_{D}(1)])\\
&  =\sqrt{2-L_{B}(\rho_{D}(1),\rho_{D}(2))^{2}}L_{B}(\rho_{D}(1),\rho
_{D}(2))\left[  \delta_{\hat{A}_D}(\rho_{D}(1))+\sqrt{\delta_{\hat{A}_D}(\rho_{2})^{2}%
+S_{D}^{2}}\right] \\
&  \leq\sqrt{2-L_{B}(\rho_{D}(1),\rho_{D}(2))^{2}}L_{B}(\rho_{D}(1),\rho
_{D}(2))\left[  \delta_{\hat{A}_D}(\rho_{D}(1))+\delta_{\hat{A}_D}(\rho_{D}(2))+S_{D}\right].
\end{align*}
Note that a similar inequality appears in \cite{11}. But the above inequality is more stringent.
This result can be rewritten as%

\[
\frac{S_{D}}{N_{D}}\leq\frac{L_{B}\left(  \rho_{D}(1),\rho_{D}(2)\right)
\sqrt{2-L_{B}\left(  \rho_{D}(1),\rho_{D}(2)\right)  ^{2}}}{1-L_{B}\left(
\rho_{D}(1),\rho_{D}(2)\right)  \sqrt{2-L_{B}\left(  \rho_{D}(1),\rho
_{D}(2)\right)  ^{2}}}.
\]
By using the relation between quantum fidelity and Bures distance 
\[
L_{B}\left(  \rho,\rho^{\prime}\right)  =\sqrt{1-F\left(  \rho,\rho^{\prime
}\right)  },
\]
we get an upper bound for the SNR as
\begin{equation}
\frac{S_{D}}{N_{D}}\leq\frac{\sqrt{1-F\left(  \rho_{D}(1),\rho_{D}(2)\right)
^{2}}}{1-\sqrt{1-F\left(  \rho_{D}(1),\rho_{D}(2)\right)  ^{2}}}. \label{33}%
\end{equation}
Note that the fidelity $F\left(  \rho,\rho^{\prime}\right)  $ obeys the
monotonicity property in any quantum channel  $\Gamma$ \cite{12}:
\[
F\left(  \Gamma\left[  \rho\right]  ,\Gamma\left[  \rho^{\prime}\right]
\right)  \geq F\left(  \rho,\rho^{\prime}\right)  .
\]
By using the monotonicity, it turns out that
\begin{equation}
F\left(  \rho_{D}(1),\rho_{D}(2)\right)  \geq F\left(  \rho_{S}(1),\rho
_{S}(2)\right)  \label{32}%
\end{equation}
holds for \ the quantum states $\rho_{S}(1),\rho_{S}(2)$ of $S$ via eq.(\ref{31}) as follows:
\begin{align*}
&  F\left(  \rho_{D}(1),\rho_{D}(2)\right) \\
&  =F\left(  \operatorname*{Tr}_{S}\left[  U\left(  \rho_{S}(1)\otimes\rho
_{D,0}\right)  U^{\dag}\right]  ,\operatorname*{Tr}_{S}\left[  U\left(
\rho_{S}(2)\otimes\rho_{D,0}\right)  U^{\dag}\right]  \right) \\
&  \geq F\left(  U\left(  \rho_{S}(1)\otimes\rho_{D,0}\right)  U^{\dag
},U\left(  \rho_{S}(2)\otimes\rho_{D,0}\right)  U^{\dag}\right) \\
&  =F\left(  \rho_{S}(1)\otimes\rho_{D,0},\rho_{S}(2)\otimes\rho
_{D,0}\right) \\
&  =F\left(  \rho_{S}(1),\rho_{S}(2)\right)  .
\end{align*}
Here we have used the fact that taking a partial trace as a quantum channel
increases the fidelity. Thus we obtain eq. (\ref{32}). Note that
$I(x)=\frac{\sqrt{1-x^{2}}}{1-\sqrt{1-x^{2}}}$ is a monotonically deceasing
function for $x\in\left[  0,1\right]  $. From eq. (\ref{33}) and eq.
(\ref{32}), our main result in eq. (\ref{3}) is derived. $\Box$

When the Hilbert space dimension of $D$ is equal to the Hilbert space dimension of $S$,  we can take a SWAP operator $U$ between $D$ and $S$ as an interaction. Then the upper bound in (\ref{5}) is attained for quantum states and hermitian operators $\rho_1$, $\rho_2$ and $\hat{A}_D$ such that  $[\rho_1,\rho_2]=0$ and the equality in eq. (\ref{eq2})  holds. In that case, the physical observable $\hat{A}_D $ which achieves the bound has a very complicated form, but in principle it is fixed by eq. (\ref{attain}). \\
\quad Here we show two  examples which attains the equality. The first example is as follows:
\begin{eqnarray}
\rho_D(1)&=&\cos^2 \theta | 0\rangle \langle 0 | +\sin^2 \theta | 1 \rangle \langle 1 |,\nonumber \\
\rho_D(2)&=& \sin^2 \theta | 0\rangle \langle 0 | +\cos^2 \theta | 1 \rangle \langle 1|\nonumber ,
\end{eqnarray}
where $|0\rangle$ and $|1\rangle$ are eigenstates of the number operator of harmonic oscillators associated with eigenvalues 0 and 1,  respectively. $\theta$ is a real parameter which satisfies $0 \leq \theta <2\pi$. We define $\omega$ as the angular frequency and $a^{\dagger}$ as creation operator and $a$ as annihilation operator. When $\hat{A_D}=\hbar \omega (a^{\dagger}a -\frac{1}{2})$, the equality of eq.(\ref{eq:eq1}) is attained. In this case, $\rho_D(1)$, $\rho_D(2)$ and $\hat{A_D}$ commute with each other.\\
\quad Next we consider more nontrivial case where two states $\rho_D(1)$ and $\rho_D(2)$ do not commute with each other: 
\begin{eqnarray}
\rho_D(1)&=& p|0 \rangle \langle 0| +(1-p) |1\rangle \langle 1| ,\nonumber \\
 \rho_D(2) &=& |+\rangle \langle+| ,\nonumber
\end{eqnarray}
where $| 0 \rangle$ and $ |1 \rangle$ are eigenstates associated with eigenvalues 1, $-1$ of the Pauli matrix $\sigma_z$ respectively and $| + \rangle=\frac{1}{\sqrt{2}}(|0\rangle + |1 \rangle)$. $p$ is a real number that satisfies $0 \leq p \leq 1$.
When $\hat{A}_D =\pm(\sigma_x-1)$ with Pauli matrix $\sigma_x$, the upper bound of eq.  (\ref{eq:eq1}) is achieved. In this case two states $\rho_D(1)$ and $\rho_D(2)$ do not commute with each other, but $\hat{A}_D$ and $\rho_D(2)$ are commutable.\\
\quad Before closing this section, we comment on the generalization of the above proof. The essence of the proof is to show the existence of two vectors $\overrightarrow{x} $ and $ \overrightarrow{y}$ in any dimensions of the system. Thus the derivation of eq. (\ref{eq:classical}) is valid even when the dimension of the system is infinity if the observable's specrum is discrete. The argument after deriving eq. (\ref{eq:classical}) is also valid as long as the quantum fidelity of the two states is well defined in the infinite dimensional systems.  For the observables with continuous or singular spectra, we  conjecture that eq. (\ref{3}) also holds. Although the proof for such observables is left for future work, it is interesting to extend our setup to general quantum detection models with arbitrary observables.

\section{Applications}

\quad In this section, we discuss applications of the universal upper bound of the quantum SNR. In subsection 3.1, we consider an example of two coherent states. In subsection 3.2, we derive a fundamental upper bound of power consumption to perform quantum switching using quantum SNR. In subsection 3.3, an application for the fidelity estimation is shown.

\subsection{Coherent state case}
\quad Let us consider the case of pure initial states of $S$:
\begin{eqnarray}
\rho_S(1)&=&|\psi(1)\rangle\langle\psi(1)|,\nonumber \\
\rho_S(2)&=&|\psi(2)\rangle\langle\psi(2)|.\nonumber
\end{eqnarray}
The upper bound in eq.(\ref{3}) becomes:
\begin{equation}
\frac{S_D}{N_D} \leq \frac{\sqrt{1-|\langle\psi(1)|\psi(2)\rangle|^2}}{1-\sqrt{1-|\langle\psi(1)|\psi(2)\rangle|^2}} \notag.
\end{equation}
Suppose that $S$ is a free quantum scalar field $\phi(t,\bm{x})$ in 3+1 dimensions:
\begin{equation}
\phi(t,\bm{x})=\int \frac{d^3 \bm{k}}{\sqrt{(2\pi)^32E_{\bm{k}}}} \{\hat{a}(\bm{k})e^{-i\bm{k}\cdot \bm{x}}+\hat{a}^{\dagger}(\bm{k})e^{i\bm{k}\cdot \bm{x}}\} \notag,
\end{equation}
where $E_{\bm{k}}=\sqrt{\bm{k}^2+m^2}$, $\hat{a}^{\dagger}(\bm{k})$ is creation operator and $\hat{a}(\bm{k})$ is  annihilation operator.
The vaccum state $\ket{0}$ is defined by $\hat{a}(\bm{k}) \ket{0} =0$.
The coherent state is given as follows:
\begin{equation}
|c\rangle=\exp\left(-\frac{1}{2}\int d^3 \bm{k} |c(\bm{k})|^2\right)\exp\left(\int d^3 \bm{k} c(\bm{k}) \hat{a}^{\dagger}(\bm{k})\right)|0\rangle,\notag
\end{equation}
where $c(\bm{k})$ is a complex function of $\bm{k}$. Let us take
\begin{eqnarray}
\rho_S(1)&=&|0\rangle\langle0|,\nonumber \\
\rho_S(2)&=&|c(\bm{k})\rangle\langle c(\bm{k})|.\nonumber
\end{eqnarray}
In this case the upper bound of the SNR is 
\begin{equation}
\frac{S_D}{N_D}\leq \frac{2\sqrt{1-\exp\left(-\int d^3 \bm{k}|c(\bm{k})|^2\right)}}{1-\sqrt{1-\exp\left(-\int d^3 \bm{k}|c(\bm{k})|^2\right)}} \label{coherent}.
\end{equation}
In quantum cryptography experiments, small amplitude coherent states with $ \int d^3 \bm{k}|c(\bm{k})|^2 $ being small are often used. Then  eq. (\ref{coherent}) provides  a severe upper bound such that
\begin{equation}
\frac{S_D}{N_D} \leq 2\sqrt{\int d^3 \bm{k}|c(\bm{k})|^2}+O\left(\left(\int d^3 \bm{k}|c(\bm{k})|^2\right)^{\frac{3}{2}}\right).
\end{equation}

\subsection{Fundamental upper  bound of power consumption to perform quantum switching}
\quad Let us consider an application of our result to derive a fundamental upper bound for the power consumption of rapid quantum switching in a short time duration $\tau$. Quantum switches consist of a control system $C$ and a target system $T$. We consider two different initial  states of $C$, $\rho_C(on)$ and $\rho_C(off)$. The initial state of $T$ is represented by $\rho_T(0)$. When the initial state of $C$ is $\rho_C(on)$,  the target system is switched from $\rho_T(0)$ to $\rho_T(\tau,on)$. Let us assume that when the initial state of $C$ is $\rho_C(off)$, the state of $T$ is unchanged. The total Hamiltonian is denoted by $H=H_T +H_C +V_{CT}$, where $H_T$ and $H_C$ are free Hamiltonians of each system and $V_{CT}$ represents the interaction between $C$ and $T$. When the initial state of $C$ is $\rho_C(on)$, the time evolution  of $T$ is given by

\begin{equation}
\rho_T(\tau,on)-\rho_T(0)=-i\tau{\rm Tr}_C[H,\rho_T(0)\otimes\rho_C(on)]+O(\tau^2).\notag
\end{equation}
The energy cost of switching is  
\begin{equation}
{\rm Tr}_T [H_T(\rho_T(\tau,on)-\rho_T(0))]=-i\tau{\rm Tr}(H_T[H,\rho_T(0)\otimes\rho_C(on)]).\notag
\end{equation}
From the cyclic rule of the trace and the property that $H_T$ commutes with $H_C$,\\
${\rm Tr}(H_T[H,\rho_T(0)\otimes\rho_C(on)])={\rm Tr}([H_T\otimes I_C,H](\rho_T(0)\otimes\rho_C(on)))={\rm Tr}([H_T\otimes I_C,V_{TC}](\rho_T(0)\otimes\rho_C(on)))$holds. The power consumption to switching is defined by
\begin{equation}
P=\frac{{\rm Tr}[H_T \rho_T(\tau,on)-\rho_T(0)]}{\tau}.\notag
\end{equation}
 Note that in the case where the initial state of $C$ is $\rho_C(off)$, the state of $T$ is unchanged, so the energy cost and the power consumption are zero. Therefore the power consumption needed to switch  becomes as follows:
\begin{eqnarray}
P &=& -i{\rm Tr}([H_T\otimes I_C,V_{TC}]\{\rho_T(0)\otimes\rho_C(on)\})\nonumber \\
&=&-i{\rm Tr}([H_T\otimes I_C,V_{TC}]\{\rho_T(0)\otimes(\rho_C(on)-\rho_C(off))\}). \nonumber
\end{eqnarray}
We remark that eq. (\ref{33}) can be rewritten as follows:
\begin{eqnarray}
& &\left\vert \mathrm{Tr}[\hat{A}_D\rho_{D}(1)]-\mathrm{Tr}[\hat
{A}_D\rho_{D}(2)]\right\vert \nonumber \\
&\leq&\frac{\sqrt{1-F(\rho_D(1),\rho_D(2))^2}}{1-\sqrt{1-F(\rho_D(1),\rho_D(2))^2}}\left[
\delta_A(\rho_D(1))+\delta_A(\rho_D(2))\right] ,\nonumber
\end{eqnarray}
where $\delta_{\hat{A}_D}(\rho)$ is the standard deviation of $\hat{A}_D$ defined  in Sec. 2.
Suppose that the system $D$ is regarded as a composite system $C+T$ and \\$\hat{A}_D = -i[H_T\otimes I_C,V_{TC}]$. We substitute  $\rho_D(1)=\rho_{C+T}(on)=\rho_T(0)\otimes \rho_C(on)$ and $\rho_D(2)= \rho_{C+T}(off)=\rho_T(0)\otimes \rho_C(off)$. Then we find 
\begin{eqnarray}
|P|&\leq&\frac{\sqrt{1-F(\rho_C(on),\rho_C(off))^2}}{1-\sqrt{1-F(\rho_C(on),\rho_C(off))^2}}  \nonumber \\
& & \times \left[
\delta_A(\rho_{C+T}(on))+\delta_A(\rho_{C+T}(off))\right]. \label{power}
\end{eqnarray}
\quad This inequality implies that the rapid quantum switching has a tight constraint from  the quantum fluctuation of the physical observable $\hat{A}_D$. Similar inequalities have been proven in \cite{13},\cite{14}. We cannot apply their results to infinite dimensional systems including harmonic oscillators since the spectrum norms of the target Hamiltonian $\|H_T\|$ may diverge.   However our result eq. (\ref{power}) is written by non-divergent quantum fluctuations. Thus, we are able to give nontrivial upper bounds for infinite dimensional systems. 

\subsection{Fidelity Estimation }
\quad An application for the fidelity estimation is also possible.
Eq. (\ref{eq:eq1}) can be rewritten as  follows :
\begin{equation}
F^2(\rho_D(1),\rho_D(2)) \leq 1-\left(\frac{S_D}{\sqrt{{\rm Tr}[\hat{A}_D^2\rho_D(1)]}+\sqrt{{\rm Tr}[\hat{A}_D^2\rho_D(2)]}}\right)^2 . \label{fidelity}
\end{equation}
Suppose we want to know an approximate value of the fidelity between two states $\rho_D(1), \rho_D(2)$.
The upper bound of the fidelity is given by the right hand side of eq. (\ref{fidelity}). We compare with the result given in \cite{15} which is:

\begin{equation}
F^2(\rho_D(1),\rho_D(2)) \leq {\rm Tr}[\rho_D(1)\rho_D(2)]+\sqrt{(1-{\rm Tr}[\rho_D(1)^2])(1-{\rm Tr}[\rho_D(2)^2])}. \label{19} 
\end{equation}
The right hand side of eq. (\ref{19}) can be fixed by performing a controlled-SWAP test. 
On the other hand, our bound given in eq. (\ref{fidelity}) is easily measurable since it can be fixed by measuring  the observable $\hat{A}_D$. It is worth stressing that our bound sometimes gives a more stringent upper bound. For example, consider the case where matrix representations of two states are 
\begin{eqnarray}
\rho_D(1) = 
\left( \begin{array}{ccc}
\frac{1}{2} & 0 & 0 \\
0 & \frac{1}{6} & 0\\
0 & 0 & \frac{1}{3}
\end{array}
\right) ,\nonumber 
\end{eqnarray}
\begin{eqnarray}
\rho_D(2) = 
\left( \begin{array}{ccc}
\frac{1}{3} & 0 & 0 \\
0 & \frac{1}{6} & 0\\
0 & 0 & \frac{1}{2}
\end{array}
\right) ,\nonumber 
\end{eqnarray}
and the observable $\hat{A}_D$ is fixed as 
\begin{eqnarray}
\hat{A}_D = 
\left( \begin{array}{ccc}
-1 & 0 & 0 \\
0 & 0 & 0\\
0 & 0 & 1
\end{array}
\right). \nonumber 
\end{eqnarray}
In this case, our bound becomes $\frac{29}{30} \simeq 0.967$. On the other hand, the bound given in \cite{15} is $\frac{35}{36} \simeq 0.972$. Therefore, our upper bound is tighter in this case.

\section{Summary}
\quad We proved the fundamental upper bound for the SNR of quantum detectors in eq. (\ref{3}) when the observable has a discrete spectrum. We conjecture that our bound also holds for observables with continuous or singular spectra, although a rigorous proof is left for future study. Our bound is computed using the information of the signal system $S$. As we have shown, this bound is independent of the interaction between $S$ and the detector system $D$ and its observable $\hat{A}_D$. Moreover, our result is more stringent than the previous result in \cite{11}. In Section 3, we have shown applications of eq. (\ref{3})  and eq. (\ref{5}). In subsection 3.1, we illustrated the upper bound of quantum SNR for coherent states which are important in quantum optics. In subsection 3.2, we derived the power consumption bound of rapid quantum switching in eq. (\ref{power}). This bound can be applied to infinite dimensional systems including harmonic oscillators and quantum fields. Finally, the application for the fidelity estimation has been provided in subsection 3.3. Our upper bound of fidelity (\ref{fidelity}) is easily measurable and sometimes becomes more stringent than previous result \cite{15}.\\
\quad  The fidelity estimation is one of applications for the state  discrimination. We can also consider other applications of eq. (\ref{3}).  One of them is weak measurements. In weak measurements, the interactions between systems and measurement devices are weak. Therefore it is difficult to discriminate system states before and after measurements. This is consistent with our bound. The fidelity between states before and after measurements is close to one and SNR becomes small. We expect that our analysis can be applied in studies on the improvability of the efficiency of state discrimination in weak measurements.         \\ \\
\quad \textit{Acknowlegement.-} We would like to thank H. Tajima, K. Saito, K.
Matsumoto, A. Kempf and E. Martin-Martinez for valuable discussion and
information. This research is partially supported by JSPS KAKENHI Grant Number
JP19K03838 (M. H.) and JP18J20057 (K. Y.), and by Graduate Program on Physics
for the Universe (GP-PU), Tohoku University (K. Y.), and by the WISE Program for
AI Electronics, Tohoku University (R. K.).


\bigskip

\begin{thebibliography}{99}                                                                                               %
\bibitem{Helstrom}
C.W. Helstrom, \textit{Quantum Detection and Estimation Theory} (Academic Press, New York, 1976).
\bibitem{Holevo}
A.S. Holevo, Jour. Multivar. Anal. 3, 337 (1973).
\bibitem{Vogel}
K. Vogel and H. Risken, Phys. Rev. A 40, 2847 (1989).
\bibitem{Leonhardt}
U. Leonhardt, \textit{Measuring the Quantum State of Light} (Cambridge Univ. Press, 1997).
\bibitem{Fujiwara2}
A. Fujiwara and H. Nagaoka, Phys. Lett. 201 A 119 (1995).
\bibitem{Matsumoto2}
K Matsumoto J. Phys. A: Math. Gen. 35, 3111 (2002).
\bibitem{Turchette}
Q. A. Turchette, C. J. Hood, W. Lange, H. Mabuchi, and H. J. Kimble
Phys. Rev. Lett. 75, 4710 (1995).
\bibitem{Chuang}
I. L. Chuang and M. A. Nielsen, J. Modern Optics, 44, (1997).
\bibitem{Poyatos}
J. F. Poyatos, J. I. Cirac, and P. Zoller Phys. Rev. Lett. 78, 390 (1997).
\bibitem{Fujiwara} 
A. Fujiwara, Phys. Rev. A 63, 042304 (2001).
\bibitem{Imai}
H. Imai and A. Fujiwara, J. Phys. A: Math. Theor. 40, 4391 (2007).
\bibitem{Amari}
S. Amari and H. Nagaoka, \textit{methods of information geometry}, (American Mathematical Society, 2000). 
\bibitem{Miller}
W. A. Miller, Proc. SPIE 10660, Quantum Information Science, Sensing, and Computation X, 106600H (2018).
\bibitem{Cafaro}
C. Cafaro and P. M. Alsing Phys. Rev. E 97. 042110 (2018).
\bibitem{Ozawa2}
M.Ozawa Phys. Rev. A 67, 042105 (2003).
\bibitem{Wigner}
E. P. Wigner, Z. Physik 133, 101 (1952).
\bibitem{Araki}
H. Araki and M. M. Yanase, Phys. Rev. 120, 622 (1960).
\bibitem{Ozawa}
M. Ozawa Phys. Rev. Lett. 89, 057902 (2002).
\bibitem{Karasawa1}
T. Karasawa and M. Ozawa, Phys. Rev. A 75, 032324 (2007).
\bibitem{Karasawa2}
T. Karasawa, J. Gea-Banacloche, M. Ozawa, J. Phys. A: math. Theor. 42, 225303 (2009).
\bibitem{Tajima}
H. Tajima, H.Nagaoka, arXiv:1909.02904.
\bibitem {1}Y. Feng, and A. I. Solomon, Opt. Commun.152, 299 (1998).

\bibitem {2}M. A. Rubin, and S. Kaushik, Opt. Lett. 32, 1369 (2007).

\bibitem {3}M. A. Rubin, and S. Kaushik, Appl. Opt. 48, 4597 (2009).

\bibitem {4}D. J. Starling, P. B. Dixon, A. N. Jordan, and J. C. Howell, Phys.
Rev. A80, 041803 (2009).

\bibitem {5}Y. Kedem, Phys. Rev. A 85, 060102 (2012).

\bibitem {6}G. C. Knee, and W. J. Munro, Rev. A 92.012130 (2015).

\bibitem{David} D. J. Starling, P. B. Dixon, A. N. Jordan, and J. C. Howell, Phys. Rev. A 80, 041803(R) (2009).

\bibitem{Lee} J. Lee and I. Tsutsui, Quantum Stud.: Math. Found. 1, 65 (2014). 

\bibitem{Susa} Y. Susa and S. Tanaka, Phys. Rev. A 92, 012112 (2015).

\bibitem {7}E. Lantz, P.-A. Moreau, and F. Devaux, Phys. Rev. A 90, 063811 (2014).

\bibitem {8}M. Reichert, H. Defienne, and J. W. Fleischer, Phys. Rev. A 98, 013841 (2018).

\bibitem {9}J. Zhao, J. Dias, J. Y. Haw, T. Symul, M Bradshaw, R. Blandino, T. Ralph, S. M. Assad, and P. K. Lam , Optica 4 [11], 1421 (2017).

\bibitem {10}G. Scala, M. D'Angelo, A. Garuccio, S. Pascazio, F. V. Pepe, Phys. Rev. A 99, 053808 (2019).

\bibitem{naimark} M. A. Naimark, Comptes Rendus (Doklady) de l'Acadenie des Sience de l'URSS, 41, 9, 359, (1943).
\bibitem{ozawa-povm} M. Ozawa, J. Math. Phys., 25, 79 (1984).
\bibitem {11}H. Tajima, N. Shiraishi and K. Saito, Phys. Rev. Lett. 121, 110403 (2018).

\bibitem{12}  M. A. Nielsen and I. L. Chuang, \textit{Quantum Computation and Quantum Information}, (Cambridge Univ. Press, 2000) 

\bibitem{13} I. Marvian, R. W. Spekkens, and P. Zanardi, Phys. Rev.
A 93, 052331 (2016).
\bibitem{14}K. Ito and T. Miyadera, arXiv:1711.02322.

\bibitem{15}
J. A. Miszczak, Z. Pucha{\l}a, P. Horodecki, A. Uhlmann and K. \.{Z}yczkowski, Quantum Information \& Computation 9, 0103 (2009).

\end{thebibliography}
\end{document}